# Designer Magnetism in High Entropy Oxides

Alessandro R. Mazza[1], Elizabeth Skoropata[1], Yogesh Sharma[1,2], Jason Lapano[1], Thomas W. Heitmann[3], Brianna L. Musico[4], Veerle Keppens[4], Zheng Gai[5], John W. Freeland[6], Timothy R. Charlton[7], Matthew J. Brahlek[1], Adriana Moreo[1,8], Elbio Dagotto[1,8], Thomas Z. Ward[1,*]

[1]Materials Science and Technology Division, Oak Ridge National Laboratory, Oak Ridge, Tennessee 37831, USA

[2]Center for Integrated Nanotechnologies, Los Alamos National Laboratory, Los Alamos, New Mexico 87545, USA

[3]University of Missouri Research Reactor, The University of Missouri, Columbia, Missouri 65211, USA

[4]Department of Materials Science and Engineering, University of Tennessee, Knoxville, Tennessee 37996-4545, USA

[5]Center for Nanophase Materials Sciences, Oak Ridge National Laboratory, Oak Ridge, Tennessee 37831, United States

[6]Advanced Photon Source, Argonne National Laboratory, Lemont, Illinois 60439, USA

[7]Neutron Science Division, Oak Ridge National Laboratory, Oak Ridge, Tennessee 37831, USA

[8]Department of Physics and Astronomy, University of Tennessee, Knoxville, TN 37996, USA

[*]email: wardtz@ornl.gov

Disorder can have a dominating influence on correlated and quantum materials leading to novel behaviors which have no clean limit counterparts. In magnetic systems, spin and exchange disorder can provide access to quantum criticality, frustration, and spin dynamics, but broad tunability of these responses and a deeper understanding of strong limit disorder is lacking. In this work, we demonstrate that high entropy oxides present a previously unexplored route to designing quantum materials in which the presence of strong local compositional disorder hosted on a positionally ordered lattice may be exploited to generate highly tunable emergent magnetic behaviors—from macroscopically ordered states to frustration-driven dynamic spin interactions. Single crystal $La(Cr_{0.2}Mn_{0.2}Fe_{0.2}Co_{0.2}Ni_{0.2})O_3$ films are used as a structurally uniform model system hosting a magnetic sublattice with massive microstate disorder in the form of site-to-site spin and exchange type inhomogeneity. A classical Heisenberg model is found to be sufficient to describe how compositionally disordered systems can paradoxically host long-range magnetic uniformity and demonstrates that balancing the populating elements based on their discrete quantum parameters can be used to give continuous control over ordering types and critical temperatures. Theory-guided experiments show that composite exchange values derived from the complex mix of microstate interactions can be used to design magnetic phase degeneracy. These predicted materials are synthesized and found to possess an incipient quantum critical point when magnetic ordering types are designed to be in direct competition; this leads to highly controllable exchange bias behaviors in the monolithic single crystal films previously accessible only in intentionally designed bilayer heterojunctions.

# I. INTRODUCTION

Magnetism is the most easily observable quantum phenomenon and is a cornerstone of technologies ranging from magnetic memory, to spintronics, to future quantum sensing and computing applications [1–4]. The type and strength of magnetic state in crystalline materials is dictated by discrete values, such as the number of unpaired electrons and type of magnetic exchange pathways populating the crystal lattice. The development of predictive design strategies aimed at tailoring functional magnetic responses is then entirely reliant on our ability to not only computationally forecast what parameters must be present on a lattice to generate a required magnetic behavior, but these parameters must also be translated into real materials through synthesis. This is fundamentally challenging, since direct continuously tunable control over spin ($S$) and magnetic exchange ($J$) values is needed to access a precisely defined parameter space; however, these values must be experimentally created using the limited set of spin active elements. There are several indirect methods used to influence magnetic responses in strongly correlated materials, such as where heteroepitaxial effects and defect engineering are commonly used to manipulate spin-coupled charge and orbital parameters [5–9]. Direct modification to the underlying $S$ and $J$ values using substitutional doping approaches are traditionally limited by thermodynamic processes during synthesis which can cause like elements to cluster or form secondary phases [10]. Thus, while substitutional doping promises the most direct route to accessing the magnetic parameters used in computational approaches, enthalpic effects during synthesis can drive element segregation which limits mixing and reduces the number of desired composite microstates that exist in well-mixed regions [11]. Recent developments in entropy-assisted synthesis provide a path to overcoming enthalpic effects and create well mixed systems [12].

By greatly increasing the number of elements present in a material during synthesis, it is possible for entropic effects to dominate during crystal formation [13,14]. The governing entropy ensures exceptional mixing, which maximizes the number of local microstates hosted on a lattice [15]. Increasing local disorder in entropy stabilized materials is linked to significant functional improvements in thermal transport [16,17], ionic conductivity [18,19], and catalytic responses [20,21] over less complex materials. Recent work has shown that this compositional disorder can be accommodated on ordered single crystalline lattices, which reduces the need to consider extrinsic parameters in computational models [22–25]. With the development of single crystal synthesis of these materials, we can now begin to more rationally design these systems to address previously inaccessible fundamental questions related to disorder while helping the applied communities disentangle intrinsic from extrinsic effects that might be present in non-single crystal form factors. In this way, the effects of known element-specific parameters and inter-atomic couplings can be microscopically mapped to materials of extraordinary compositional complexity to predict macroscopic collective behavior, and critically, these compositionally complex but structurally uniform crystals can be synthesized in the real world.

Unlike high entropy alloys built from metal-metal bonded elements [13,26], high entropy oxides give access to functionalities in covalent and ionic bonded materials. While the metal-bonded high entropy alloys are limited in their range of crystal structures, stability, and accessible magnetic interaction pallet, the addition of an anion sublattice enables a broad range of stable crystal structures and greater access to functional diversity [14,15]. Still, considering the sensitivity of spin behavior to bond angles and cation orbital filling in many oxide systems [27,28], it is surprising that high entropy oxides [29] hosting high levels of compositional disorder have been reported to support signatures of long range magnetic ordering in both relatively simple rock salt lattices [30–32] and more complicated spinel [33,34] and perovskite [35,36] structures. Of broader interests, the local exchange and spin disorder hosted in these compositionally complex systems may provide critical insights on the role of disorder in degeneracy-driven magnetic dynamics and phase competition. In all cases, the mechanism of ordering must revolve around the type and strength of exchange couplings populating the lattices as a function of the elemental compositions. This is fundamentally different from the itinerant RKKY or dilute type magnetism observed in metal-metal bonded systems [13,37]. The addition of the anion sublattice in the high entropy oxides promises a greater diversity of magnetic interactions enabled by localized rather than delocalized electron coordination. However, the presence of correlated states in high entropy transition metal oxides and the resulting increase in complex microstates comes at the cost of greatly decreasing the feasibility of producing accurate descriptions of these systems using computationally intensive first principles approaches [38].

In this work, a range of single crystal entropy-stabilized $AB$O$_3$ perovskite films are synthesized to probe how manipulating variances in site-to-site spin and exchange interactions influence magnetic behaviors. In spite of the extraordinary levels of microstate complexity, a classical Heisenberg model is found to be capable of producing an astonishingly accurate magnetic phase diagram. This computationally streamlined approach provides a useful framework to understand how long-range antiferromagnetic order can paradoxically emerge from the disordered chaotically dispersed coupling parameters in correlated high entropy oxides. The model additionally provides a practical means of predicting how to manipulate the parameter ratios to also stabilize ferromagnetism or drive degeneracy-induced metastability. A series of theory-guided experiments confirm the viability of the model and show that manipulating the *S* and *J* parameters through compositional ratio allows access to an incipient quantum critical point (QCP) at an AFM to FM phase crossover, where the phases are shown to be in direct competition. We also discover that manipulating the local spin disorder can be used to drive exchange bias behaviors in the monolithic single crystal films similar to that observed in AFM-FM bilayer heterojunctions; this provides important new insights into recent proposals that spin and exchange disorder can be the dominating factor in generating exchange bias responses [39].

## II. RESULTS

La($Cr_{0.2}Mn_{0.2}Fe_{0.2}Co_{0.2}Ni_{0.2}$)$O_3$ (L5BO) is an ideal high entropy oxide system to examine the role of local magnetic disorder on the emergence of macroscopic magnetic behaviors. Structurally, this *AB*$O_3$ perovskite possesses full mixing of the *B*-site cations while maintaining long-range single crystal lattice uniformity [25,36]. There is a large body of theoretical and experimental work on the parent compounds, $LaCrO_3$, $LaMnO_3$, $LaFeO_3$, $LaCoO_3$, and $LaNiO_3$ and a range of interfacial and co-doping studies that provide insights into expected spin, charge, and oxygen mediated coupling types between different *3d* transition metal cations across a range of structural distortions and dimensionalities [40–57]. However, even with this large body of work to draw upon, functional theoretical modelling of the high entropy oxides has been extremely limited. Electron correlation and many-body effects combined with the range of possible microstates in a compositionally complex system make precise quantitative calculations impractical [58]. Since the importance of a single microstate is reduced as the number of possible nearest neighbor combinations increases, it is computationally unrealistic to focus on perfect reproduction of all states in the lattice [59]. For example, a *B*-site in a simple *AB*$O_3$ perovskite is surrounded by like elements on the *B*-O-*B* sublattice so has only 1 possible nearest neighbor arrangement; in the L5BO system, there are more than a 200 possible nearest neighbor combinations for each of the 5 different *B*-site elements.

L5BO's structural uniformity makes it well suited to allow its complex magnetic interactions to be modeled classically for a square lattice by considering the range of possible exchange interactions distributed across the lattice [60–62]. This generates a diverse and complicated array of possible exchange coupling terms and local spin environments. A calculation of the magnetic order parameter can be found by considering all possible microstates in a random distribution on the perovskite lattice. For the theoretical prediction of the magnetic behavior corresponding to L5BO, a Monte Carlo study is performed on a cubic lattice using the classical Heisenberg model defined as:

$$H = \sum_{<ij>} J_{ij} \mathbf{S}_i \cdot \mathbf{S}_j$$

$S_i$ are classical spins of different magnitudes depending on which transition metal element is placed at site *i*. The symbol *<ij>* refers to nearest-neighbor (NN) sites. For the location of the spins, a random distribution based on a probability is used such that each element covers 20% of the finite clusters employed for the simulation [59].

The fact that the cations in the L5BO system randomly populate a lattice that, while uniform, is neither identical to a specific parent material nor an average of all parents [25] makes direct harvesting of *S* and *J* values from literature difficult. There is little known about the way many of these cations will couple for a single isolated bond while neglecting the other 5 nearest neighbors, which could influence charge and orbital state. It is important to assign values that would have the highest probability of being most valid when randomly distributed throughout a chaotic compositional landscape hosted on a well-ordered lattice. In a randomly mixed system, the central element is most probably coordinated to several different transition metals. Harvesting individual parameters directly from a single source in literature without considering structural similarities to L5BO or the influence of charge renormalizations is unlikely to provide the best fit for the high entropy system.

There are uncertainties in selecting spin values in a random *B*-site occupancy, which results from the influence of nearest neighbors that are not present if these values were to be taken from previous reports on low complexity ternary or quaternary parent systems (Fig. 1). Charge redistribution and disproportionation can occur when linking certain transition metals through an oxygen bond which changes accessible charge state and resulting spin state; the most well-known being the strong preference of $Mn^{3+} + Ni^{3+} \rightarrow Mn^{4+} + Ni^{2+}$ and the slightly less favorable $Mn^{3+} + Co^{3+} \rightarrow Mn^{4+} + Co^{2+}$ [51,63]. This charge rebalancing is a known process in L5BO crystals as a driving factor in generating its unexpected crystalline uniformity and must be considered when selecting parameter values that have the highest probability of being in the majority [25]. With these considerations, the magnitudes of the spin of each transition metal ion are $S=5/2$ for Fe, $S=2$ for Co, $S=3/2$ for Mn and Cr, and $S=1$ for Ni [59]. For the theory results described below, one of the key factors determining the overall dominance of AFM is that $Fe^{3+}$ has antiferromagnetic tendencies in 4 of its 5 bonds *i.e.,* in Fe-O-Fe, Fe-O-Cr, Fe-O-Co, and Fe-O-Ni, where O is the bond oxygen. $Fe^{3+}$ is assigned a value $S=5/2$ because it corresponds to one electron per *3d* orbital ($3d^5$) which maximizes the spin due to the strong Hund coupling of Fe.

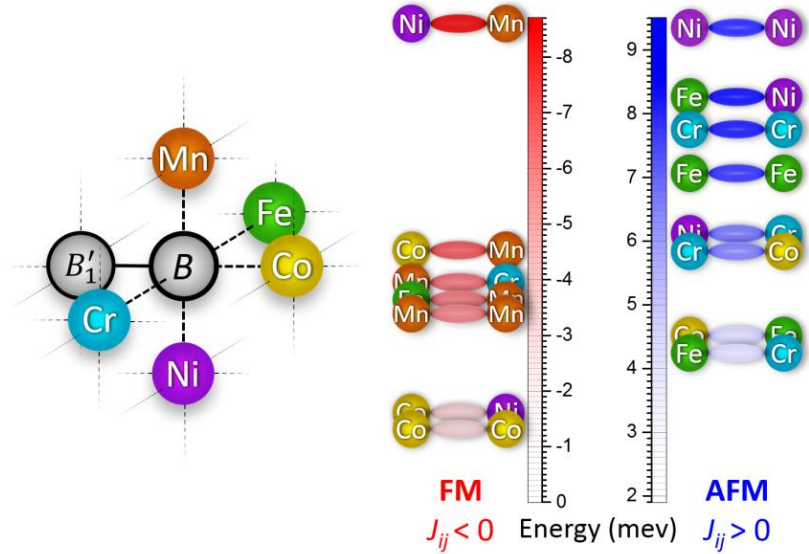

FIG. 1. Overview of complex coordination and exchange parameters. (Left) Correlation and charge compensation driven by the complexity of the 6-fold coordinated nearest neighbors around a central *B*-site can influence *B*-O-*B'* interaction. (Right) Identifying the magnetic exchange (*J*) values rely on a holistic investigation of literature to best match the highest probability of state to the known L5BO structural parameters. There are 15 types of magnetic exchange interactions extracted and inferred from literature used to model L5BO. These interactions range from ferromagnetic (FM) to antiferromagnetic (AFM). See supplementary materials for details.

The spin values are relatively easy to predict in magnitude once charge rebalancing is considered, but the values $J_{ij}$ are more difficult because there are 15 possible combinations X-O-Y (with X,Y = Cr, Mn, Fe, Co, Ni). A detailed analysis along with the existing experimental literature used to deduce $J_{ij}$ can be found in supplemental material [59]. For the 5 X-O-X cases, this task is simplified, because the critical temperatures of the $LaXO_3$ materials can be directly related to their superexchange $J$, positive or negative depending on whether the order is AFM or FM. For the other 10 X-O-Y (X≠Y) cases, finding $J_{ij}$ is more challenging and critical temperatures for 50% mixes $LaX_{0.5}Y_{0.5}O_3$, superlattices $LaXO_3$-$LaYO_3$, or in some cases interpolations between existing data are used. The summary of this effort is the following. There are 8 AFM $J_{ij}>0$ and 7 FM $J_{ij}<0$ commensurate with Fig. 1, suggesting a fine balance between the two tendencies. However, the actual magnitudes of the AFM set $J_{ij}>0$ are considerably stronger than those with FM character. Moreover, as explained above, the largest spin $S=5/2$ for $Fe^{3+}$ plays an important role for the AFM dominance. The FM tendencies are primarily driven by the links containing Mn.

The results shown in Fig. 2 use 10×10×10 clusters, with periodic boundary conditions in all directions, and the Monte Carlo simulation technique to model expected magnetic responses of the L5BO system and two related and well-studied ternary systems. An annealing process from high temperature (i.e., slow cooling) is employed to avoid being trapped into metastable states. Some of the simulations are repeated using 12×12×12 clusters and the results for critical temperatures do not change appreciably considering the uncertainties in superexchange values used. Moreover, the Monte Carlo results shown correspond to averages over 5 independent random distributions of spins, but self-averaging renders the 5 results nearly identical within the accuracy needed. This comparison provides an interesting point, by suggesting it is this average which dictates the predominant order type in the film. After equilibrium, Monte Carlo is used to measure the standard spin-spin correlations $<S_i \cdot S_i>$ in real space at all distances and from them calculate the spin structure factor $S(\boldsymbol{k})$ by Fourier transform:

$$S(\boldsymbol{k}) = \frac{1}{N}\sum_{i,j}\langle S_i \cdot S_j\rangle e^{i\boldsymbol{k}\cdot(\boldsymbol{r}_i-\boldsymbol{r}_j)}$$

$\boldsymbol{r}_i$ is the vector position in the cubic lattice of site $i$. Among allowed momenta, we searched for the one that maximized $S(\boldsymbol{k})$. In all simulations of the equiatomic $B$-site perovskite using the couplings and spins discussed above, the dominant peak in $S(\boldsymbol{k})$ is always found to be located at $(\pi, \pi, \pi)$, i.e., in the AFM position. Despite the many FM couplings used, no peak at $(0,0,0)$ is detected. Typical results are presented in Fig. 2 where $S(k_{AFM})$, with $k_{AFM} = (\pi, \pi, \pi)$, is shown as a function of temperature for the pure 100% Fe-O-Fe case (with the largest spin), the 100% Ni-O-Ni (with the smallest spin), and the 20% equiatomic $B$-site populated L5BO. Considering that $T_N = 740$ K for pure LaFeO$_3$, this establishes the scale $J$ in Fig. 2 to be $\approx 82$ K after rescaling of existing Monte Carlo results, giving a theoretical prediction that L5BO is AFM with $T_N \approx 210$ K [64].

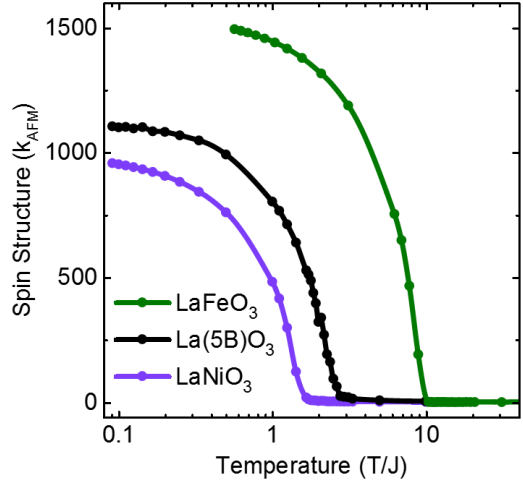

FIG. 2. Comparison of calculated spin structure factor $S(k_{AFM})$ at wavevector $k_{AFM} = (\pi, \pi, \pi)$ vs temperature, $T/J$, where $J$ is taken as 82 K, for LaFeO$_3$, LaNiO$_3$, and entropy-stabilized La(Cr$_{0.2}$Mn$_{0.2}$Fe$_{0.2}$Co$_{0.2}$Ni$_{0.2}$)O$_3$. The model matches known transition behaviors for the simple ternary compounds and predicts a G-type antiferromagnetic ordering in the L5BO system with a $T_N \approx 210$ K.

Experimentally, neutron diffraction confirms that this complex mix of local microstates creates long-range magnetic ordering in both the bulk powder ceramic and single crystal thin film forms (Fig. 3(a)-(b)). These results are presented considering the cubic Miller indices of the film, where the (0 0 1) peak refers to the structural and temperature-independent feature in the powder diffraction, and the (½ ½ ½) peak is an AFM Bragg peak occurring only below $T_N$. The temperature dependence of the (½ ½ ½) peak of the L5BO in polycrystalline form provides an order parameter that shows an onset of AFM occurring between the measurements taken at 300 K and 150 K, which is consistent with Mössbauer studies in previous studies of L5BO powder produced by spray pyrolysis [35]. Since single crystals may help preclude possible extrinsic contributions related to complexity of grain size, surface effects, and inhomogeneous mixing of constituents, neutron diffraction is also performed on single crystal films grown on near lattice matched (LaAlO$_3$)$_{0.3}$(Sr$_2$TaAlO$_6$)$_{0.7}$ (LSAT) substrates to allow the film to maintain a nearly cubic structure such as that used in the Monte Carlo simulations [36,59]. The temperature-dependent evolution of the (½ ½ ½) peak for a 90nm film is shown in Fig. 3(b) and demonstrates a clear G-type AFM transition in the L5BO film. These single crystal films show no sign of relaxation and have rocking curve widths <0.08°. While the exact onset temperature is somewhat hidden by the substrate background signal at higher temperatures, the onset trend agrees with irreversibility in temperature-dependent magnetization in the field cooled vs zero field cooled SQUID magnetometry results, implying that the Neel temperature occurs near 180 K [59].

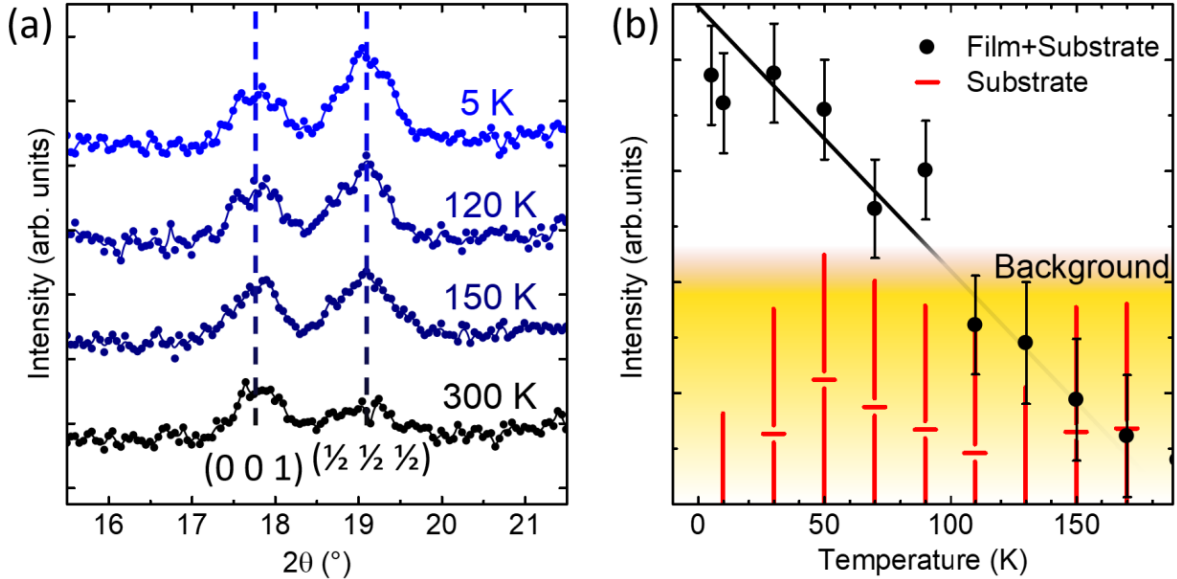

FIG. 3. Observation of G-type antiferromagnetic ordering in L5BO bulk and film. (a) Temperature dependent neutron diffraction of bulk polycrystalline L5BO sample shows the (½ ½ ½) peak emerge and increase in intensity with decreasing temperature which is consistent with the onset G-type antiferromagnetism. No other temperature dependent peaks were observed and the structural (0 0 1) peak is included for reference. (b) Neutron diffraction data taken for the order parameter of the (½ ½ ½) peak in a L5BO single crystal film on an LSAT substrate shows the onset of magnetic order emerging near the predicted Neel temperature.

The AFM found in simulation appears percolated rather than isolated, which matches the long-range AFM order observed in experiment. Only a single transition temperature is observed at $T_N \approx 180$ K, which remarkably matches the streamlined classical Heisenberg model's prediction. This shows that it is possible to design critical temperature and macroscopic ordering type by intelligently selecting a set of elements based on the average of their fundamental $S$ and $J$ into account. This gives access to continuously tunable magnetic phase spaces that are not accessible using less complex compositions. However, the uniformly percolated AFM state matching the Heisenberg model in the case of an average spin or exchange interaction over the whole lattice described above gives the false impression of being a "standard" AFM from the macroscopic perspective. In the equiatomic L5BO system, the FM bonds account for 40% of the total exchange interaction strength. This means that the AFM order dominates the FM order only by a narrow margin. This leaves the questions as to why FM is not more strongly observed considering that there are so many FM links. To understand how local spin and exchange disorder lead to this behavior, it is informative to consider the microscopic landscape by considering the local distribution of FM and AFM connections within the 10×10×10 simulation.

Fig. 4(a) presents a section of a typical 10×10 slice of the simulation showing the sign of the thermalized NN spin-spin correlations at each link after equilibrium. Here, regions of ferromagnetically coupled neighbors are clearly observed embedded in a continuous antiferromagnetic matrix. These FM regions appear at the same critical temperature as AFM order but are not coherently coupled among each other, which prevents percolation of the FM state. This intrinsic frustration in the L5BO magnetic lattice is highlighted by the non-collinearity of the equilibrated spins; and while AFM order dominates, the average of the parent oxides does not complete the story. When considering the distribution of superexchanges prior to equilibrium or considering only the inputs for nearest neighbor interactions produced from literature $J(S_i \cdot S_j)$, this imbalance shows that at higher temperature one might expect the FM character of the crystal to play more of a visible role [59]. As the zero-energy state is approached upon cooling, the AFM becomes much more dominant. That means that there should exist some relatively small external perturbation strength, or small energy scale, above which the FM phase may percolate within the AFM matrix and below which the FM regions remain locally isolated. The AFM state is thus predicted to be fragile, and at the verge of switching to a ferromagnet. Experimentally, this can be tested by applying a magnetic field strong enough to saturate the FM regions and then field cooling to a temperature below the blocking temperature ($T_B$) of the AFM region [39,65]. This effectively locks-in pockets of aligned moments and creates a surplus magnetization that is observable as a vertical shift in any field-dependent magnetization loops [66]. This vertical shift results from pinning of the uncompensated and FM spins and is distinctly different from horizontal loop shifts associated with magnetic exchange bias resulting from coupled AFM-FM systems where a fully formed FM component can overcome the interfacial energy and rotate fully [67,68]. Fig. 4(b) shows examples of field-dependent magnetization loops at different temperatures for a 62 nm film grown fully tensile strained to $SrTiO_3$ (STO) after field cooling under +/-7 T and are summarized in Fig. 4(c). As temperature decreases, the coercive field drops. This indicates the FM clusters effectively behave as single domains rotating freely with field as in a superparamagnetic regime–suggesting the unpinned moments are isolated at low temperature. The drop in coercivity coincides with the pinning of moments, which fits with the scenario in which the AFM phase hardens at low temperature—passing through $T_B$ and pinning uncompensated moments while isolating FM clusters. The results allow us to place a rough energy value of $E = k_B(10 K) = 0.86$ meV as the crossover above which the FM fluctuations are active.

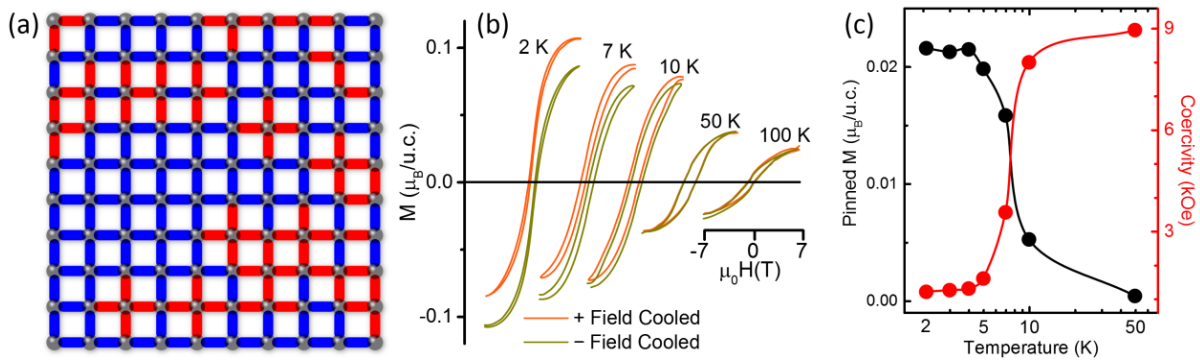

FIG. 4. Influence of frustrated and unpercolated ferromagnetic coupling. (a) Typical Monte Carlo simulation equilibrated snapshot of a 10×10 slice from the 10×10×10 cube of L5BO system show that unpercolated FM clusters can persist in a percolated AFM matrix at low temperature. Red indicates links with a positive nearest neighbor $S_i \cdot S_j$ which are mainly FM links in equilibrium low temperature while blue represent AFM linkages. (b) Field biased magnetization loops taken across a range of temperatures after field cooling under +/- 7T, show the pinned FM moments driving a vertical shift in the field biased loops below the blocking temperature, $T_B$, consistent with transition to a low temperature AFM phase populated by scattered unpercolated FM exchange sites. (c) Summary of experimentally obtained magnetization behaviors of L5BO film subjected to field cooling from 300 K demonstrates the manifestation of low temperature hindered percolation of the FM clusters. Magnetic pinning increases and coercivity decreases as the AFM phase hardens through $T_B$.

  These observations show an extraordinary correspondence between what is theoretically predicted and experimentally observed across multiple length scales. To further test this, we model the expected effects of iteratively shifting the composite state to lower *J* and perturb *S* values on the disordered lattice by increasing the ratio of Mn concentration in the lattice [59]. Analyzing the set of superexchange values used, X-O-Y links containing Mn favor ferromagnetism. Consequently, it is natural to imagine that increasing the relative Mn concentration in the Monte Carlo simulations should eventually lead to global ferromagnetism. Comparative Monte Carlo simulations provide expected spin structure factors as the percentage of Mn increases in relation to the other transition metals populating the lattice, where 20% is equiatomic L5BO (Fig. 5(a)). As % Mn initially increases to 30%, the Néel temperature decreases, but still S(0,0,0) is negligible and indicates no clear FM order. However, at 40%, the FM signal at $k_{FM}$= (0,0,0) starts developing and becomes very similar to the AFM signal at $k_{AFM}$= ($\pi, \pi, \pi$). This is indicative of a degenerate tipping point between phase preferences. Both signals grow at the same temperature upon cooling. Snapshots of MC simulations visually suggest that the FM clusters have percolated at the 40% Mn concentration, with both the FM and AFM regions being coherent over long distances. At 50% Mn or larger, the AFM $S(k)$ becomes negligible. Increasing Mn to 50% and beyond leads to the FM and AFM states switching roles, with unpercolated AFM clusters being embedded in the fully percolated FM matrix. From this, we see that it is possible to select and control the dominant percolated magnetic phase, the $T_N$ and $T_C$ of the percolated phases, and that it is even possible to balance the $J$ and $S$ composition in such a way that both AFM and FM can be fully expressed and coexisting as seen in the 40% Mn composition. This can be visualized in example cross sections from the $J(\mathbf{S}_i \cdot \mathbf{S}_j)$ 10×10 cluster matrices for three different compositions which show the evolution of the superexchange environment moving from percolated AFM at 20% Mn (Fig. 5(b)), to coexisting AFM and FM at 40% Mn (Fig. 5(c)), to percolated FM at 60% Mn (Fig. 5(d)). From these models, a theoretically predicted magnetic phase diagram is compiled (Fig. 5(e)).

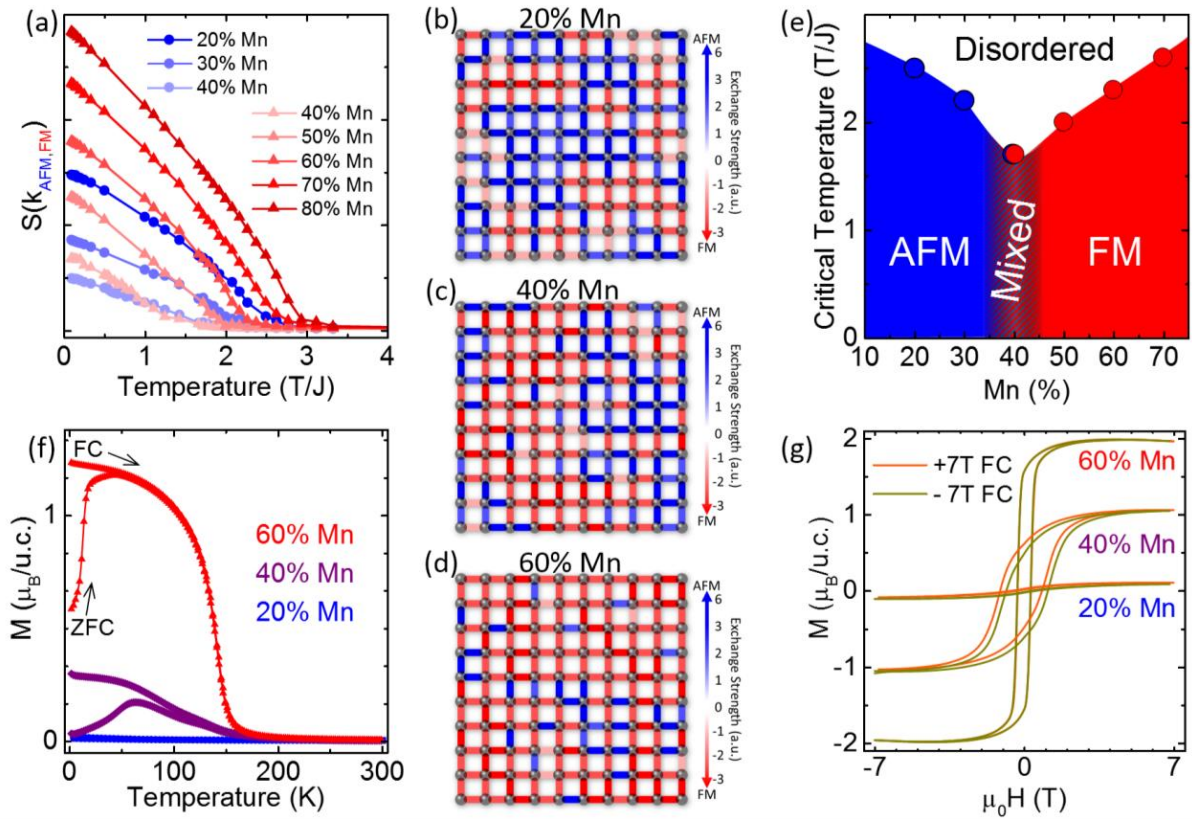

FIG. 5. Predictive modelling and experimental validation of magnetic phase diagram as a function of increasing Mn concentration. (a) Spin structure factor $S(k_{max})$ calculated by varying the % of Mn with the other 4 elements equally distributed in %, where $k_{max}$ is either (0,0,0) (FM, red) or (π,π,π) (AFM, blue) and obtained using Monte Carlo on a 10x10x10 cluster. At 40% Mn, both AFM and FM orders are percolated in the spin structure, which is why both order parameters are presented in the plot. (b)-(d) Snapshots of calculated ($J\,S_i \cdot S_j$) cross-sections for Mn concentrations corresponding to AFM, mixed AFM-FM crossover, and FM regimes show how local AFM (blue), FM (red), and near degenerate(lightly shaded) superexchange values evolve with Mn concentration. (e) The phase diagram derived from the computational model as a function of Mn content. (f) Field cooled and zero field cooled temperature dependent magnetization taken under 1 kOe field for experimentally synthesized films. (g) Magnetization loops taken at 2 K after field cooling under +/- 7 T for each of the compositions shows evolution from AFM dominated vertical offset at 20% Mn, to a horizontal exchange bias offset driven by the coexisting percolated AFM/FM phases at 40% Mn, to a FM dominated state at 60% Mn.

Single crystal films of La(Cr$_{0.15}$Mn$_{0.4}$Fe$_{0.15}$Co$_{0.15}$Ni$_{0.15}$)O$_3$ (40% Mn) and La(Cr$_{0.1}$Mn$_{0.6}$Fe$_{0.1}$Co$_{0.1}$Ni$_{0.1}$)O$_3$ (60% Mn) are synthesized to test these predictions experimentally [59]. In Fig. 5(f), temperature-dependent magnetizations of the synthesized films show that increasing Mn concentration significantly changes the magnetic responses as predicted. At the 60% Mn concentration, the 58 nm film grown on STO presents a single sharp upturn in moment consistent with a FM $T_C \approx 160$ K. The 40% Mn composition, the 56 nm film grown on STO has a slightly lower critical temperature than the 20% and 60% compositions and a more gradual transition region as would be expected for competing AFM and FM phases. If the local spin and exchange landscape in the synthesized films does mimic the complexity predicted in the model, there should be a high degree of exchange frustration and spin disorder populating these lattices. This would lead to behaviors which may mimic those traditionally observed at the interface of thin film heterostructures comprised of interfacing AFM and FM layers, where intentional design of these exchange interaction discontinuities are used to manipulate exchange bias effects for spin valve, logic, and storage applications [69]. Recent work has shown that increasing the disorder and uncompensated spin populations at the interface of these heterostructures may be the dominating influence in all AFM-FM exchange bias devices [39]. The expectation may then be that the level of spin disorder hosted in the configurationally complex high entropy oxide films should provide outstanding tunability over exchange bias responses even in single crystal monolithic systems.

Exchange bias characterization of the films shows a clear difference in response for each composition (Fig. 5(g)). To understand how these behaviors arise, it is important to consider the shifts in dominant phase energies at each composition. The energy scales involved in the magnetic response of the film are the effective Zeeman energy of the FM ($E_F$), the anisotropy energy of the AFM ($E_{AF}$), and the exchange energy at the interface ($E_I$) between the FM and AFM components [65]. For a vertical shift, such as that observed in the 20% Mn, to occur from the pinning of FM, it must be that $E_F < E_I$ and $E_F < E_{AF}$, which points to the AFM dominating the magnetic behavior and pinning the FM moments in the film as described above. With increasing

Mn concentration, the vertical loop shift observed in the 20% Mn film transitions to a horizontally shifted loop in the 40% Mn film, and finally to a state where field cooling has no obvious effect in the 60% Mn film. Given the observed shift from vertical to horizontal exchange bias response with Mn doping, it is clear that the energy scales shift relative to one another. Horizontal exchange bias implies both $E_I < E_F$ and $E_I < E_{AF}$ and that $E_F$ is strong enough to overcome the exchange energy of the interface and rotate in the biased state [67]. The horizontal shift observed in the 40% Mn film is clear evidence of a traditional exchange biasing phenomenon associated with coexisting and coupled FM and AFM phases of similar energy scale as would be expected in a system with coexisting percolated FM and AFM phases such as that predicted by theory. This control over magnetic coupling in monolithic single phase, single crystal films is remarkable, as the exchange biasing response is usually associated with heterostructured or nanocomposite magnetic materials, where direct coupling is subject to multiple crystalline components. These observations may present an important new direction in understanding the dominating mechanism of exchange bias behaviors more generally. Here, we see that manipulating the local spin disorder can be used to drive exchange bias behaviors in the monolithic single crystal films which resemble responses normally only accessible through intentionally designed heterojunctions; this provides important new insights into recent proposals that spin and exchange disorder can be the dominating factor in generating exchange bias responses [39].This suggests that the translational symmetry breaking of AFM and FM bonds gives rise to the direct coupling observed in the 40% Mn sample—similarly to that produced in artificial magnetic coupling at heterointerfaces. In the 60% Mn film, there is no longer a percolated AFM phase to bias the dominant FM phase, thus there is no observable loop shift [66]. The unpercolated small FM regions grow and begin to percolate near the region of 40% composition and come to fully dominate for the 60% composition and is also consistent with the observation that the saturation moment approaches the maximum allowed for Mn in $S = 3/2$ state.

### III.  DISCUSSION AND CONCLUSIONS

Traditional enthalpy-driven synthesis approaches often create materials that possess unintended secondary crystal phase formations or defects which generate extrinsic contributions when more than a few elements are combined. This limits our ability to cover a continuously tunable magnetic parameter phase space that can be simply modelled using only intrinsic parameter variables. Experimental access to narrow regions of calculated parameter space is a critical need to enable computational materials design strategies. The presented work demonstrates that the compositionally disordered but positionally ordered lattices produced using entropy-assisted synthesis approaches may greatly simplify our ability to create materials with very specific macroscopic magnetic responses. A classical Heisenberg model using only the intrinsic spin and exchange parameters is shown to be sufficient to predict the emergence of complex magnetic behaviors and critical temperatures in high entropy oxides comprised of many different magnetically active *3d* transition metals. Focusing on the transition metal perovskite oxides, we find that the magnetic order parameter can be predicted and synthesized with prediction and experiment matching exceptionally well. Modelling is used to direct elemental compositional ratios required to modify macroscopic expressions of AFM and FM. Shifting composite exchange values through element selection allows us to realize fully percolated AFM and FM order; but also provides guidance to the synthesis of coexisting AFM and FM phases. The ability to design unusual or frustrated competition between two or more states holds a great deal of promise for future exploration.

Manipulating frustration and local degeneracy may open new possibilities for more complex functionalities. As an example, magnetic frustration has been explored extensively on triangular, pyrochlore, and artificial lattices, where observation of dynamic magnetic behaviors such as spin liquids are generally attributed to degenerate ground states relying on geometric frustration. [70] In magnetically complex high entropy oxides such as those described in this work, it may be possible to replace geometric frustration by exchange frustration controlled by modifying the variance of exchange couplings populating the square lattice [71–73]. Further, the ability to shift local variances in spin and coupling types while maintaining position symmetries may lead to exciting opportunities for designing novel Griffiths phases or quantum many-body systems with tunable critical behaviors [74].

Including quantum fluctuations in the theoretical analysis instead of the classical spins used in the current model, may identify exactly balanced parameters leading to a quantum critical point or, alternatively, to a region of low temperature glassy behavior. Moreover, it is well known that disorder and site-to-site variance in spin and exchange parameters play an important role in nucleating regions of one characteristic or another [64,75,76]. Thus, it is reasonable to predict that manipulating variance while maintaining macroscopic averages may allow for tuning local behaviors that could drive unexpected emergent properties. As an example, considering that the oxides with only two transition metal elements can develop non-collinear states, the true phase diagrams of the compositionally complex systems likely hide more complicated responses than our simple model is able to predict. In a fully percolated phase, controlling the strength and number of randomly distributed frustrating sites into the matrix should reduce the ease with which percolation can occur, and may be utilized to provide exceptional control over transition temperatures. Manipulating the strength and type of coexisting and randomly distributed microstates populating a well-ordered single crystal offers opportunities to explore the effects of disorder in the strong limit; this is particularly important in a system where frustration and degeneracy may lead to unexpected or previously inaccessible phase spaces [77]. By mixing elements having similar magnitude of exchange strength but opposite sign, it may be possible to intentionally design metastability, dynamic responses, and near global frustration into the lattice. With large difference in the site-to-site $S$ and $J$ values, the disorder itself might become a new order parameter as it drives local phase interactions and boundary conditions.

## Acknowledgments

Experiment design, sample synthesis, structural characterization, and computational modelling were supported by the US Department of Energy (DOE), Office of Basic Energy Sciences (BES), Materials Sciences and Engineering Division. BLM thanks the Center for Materials Processing, a Center of Excellence at the University of Tennessee, Knoxville funded by the Tennessee Higher Education Commission (THEC), for financial support. This research used resources of the Advanced Photon Source, a U.S. Department of Energy (DOE) Office of Science User Facility operated for the DOE Office of Science by Argonne National Laboratory under Contract No. DE-AC02-06CH11357. A portion of the SQUID magnetometry was performed as a user project at the Center for Nanophase Materials Sciences, which is sponsored at Oak Ridge National Laboratory (ORNL) by the Scientific User Facilities Division, BES, DOE. Some neutron diffraction experiments were conducted at the Spallation Neutron Source, a DOE Office of Science User Facility operated by the Oak Ridge National Laboratory.